\documentclass[aps,prl,preprint,groupedaddress]{revtex4-2}
\usepackage{amsmath}
\usepackage{amssymb}
\usepackage{graphicx}
\usepackage{float}
\usepackage{natbib}
\usepackage[utf8]{inputenc}
\usepackage{soul}
\usepackage{caption}
\DeclareCaptionLabelSeparator{bar}{\ \textbar\ }
\begin{document}

\title{Unveiling orbital optical chirality through multipolar chiral light-matter interaction}

\author{Shun Hashiyada}
\email{shun.hashiyada@es.hokudai.ac.jp}
\affiliation{Laboratory of Nanosystem Optical Manipulation, Department of Photonics and Optical Science, Research Institute for Electronic Science, Hokkaido University, Nishi 10, Kita 21, Kita-ku, Sapporo, Hokkaido 001-0021, Japan}

\author{An'an Wu}
\affiliation{Institute of Industrial Science, University of Tokyo, 4-6-1 Komaba, Meguro-ku, Tokyo 153-8505, Japan}

\author{Yoshito Y. Tanaka}
\email{ytanaka@es.hokudai.ac.jp}
\affiliation{Laboratory of Nanosystem Optical Manipulation, Department of Photonics and Optical Science, Research Institute for Electronic Science, Hokkaido University, Nishi 10, Kita 21, Kita-ku, Sapporo, Hokkaido 001-0021, Japan}
\affiliation{Institute of Industrial Science, University of Tokyo, 4-6-1 Komaba, Meguro-ku, Tokyo 153-8505, Japan}

\date{\today}

\begin{abstract}
Chiral light–matter interactions have traditionally been understood in terms of electric–magnetic dipolar interference driven by light with spin angular momentum.  Here, we show that optical chirality can also originate from the orbital angular momentum (OAM) of light, giving rise to higher-order multipolar chiral responses.  Using a twisted gold nanorod dimer and tightly focused circularly polarized optical vortex beams carrying spin and orbital angular momenta of the same sign, we measure spectrally and spatially resolved chiral dichroism signals that persist even where spin optical chirality vanishes, revealing a quadrupole-mediated chiral interaction driven by OAM.  The spectra reveal clear quadrupole resonances whose spectral profile is strongly modulated by the OAM sign, demonstrating an OAM-driven chiral interaction.  Crucially, the signal satisfies optical reciprocity, ruling out artefacts from anisotropy or misalignment and confirming its nature as a true chiral response.  Angular momentum dissipation analysis further shows that orbital contributions dominate over spin.  These findings establish the existence of a distinct form of optical chirality, referred to as orbital optical chirality, which opens new avenues for probing and controlling multipolar chiral light–matter interactions beyond the dipolar paradigm.
\end{abstract}

\maketitle

\section{ MAIN TEXT}
Chiral light–matter interactions are central to fundamental questions of symmetry and conservation laws in electromagnetism \cite{ref1, ref2}.  These interactions also underpin major advances in stereochemistry, bio-sensing, and nanophotonics \cite{ref3}.  Among them, circular dichroism (CD), the differential absorption of left- (LH) and right-handed (RH) circularly polarized (CP) light, which carries spin angular momentum (SAM) \cite{ref4, ref5}, has long served as a canonical probe of molecular and material chirality \cite{ref6, ref7}.  In isotropic chiral media, CD arises from electric and magnetic dipole transitions and can be described as the product of two chiral quantities: the material’s chiral susceptibility and the optical chirality (OC) of the electromagnetic field, as formulated by Tang and Cohen \cite{ref8}.  This framework clarified the physical significance of OC, a conserved quantity whose flux corresponds to optical SAM. We refer to this SAM-derived field chirality as \textit{spin} OC.

Because the SAM associated with CP light is defined at a single point in space \cite{ref9}, CP light is naturally suited to coupling with point-like dipolar transitions.  However, it lacks the spatial extent necessary to efficiently excite higher-order multipolar modes, which involve spatially distributed current or charge configurations.  As a result, CD measurements using CP light are effectively limited to dipolar responses in isotropic media.  Dipolar responses alone cannot fully capture the rich modal structure of complex chiral materials.  To overcome this limitation, structured light carrying orbital angular momentum (OAM) has been proposed as an alternative probe.  Optical vortices (OVs) with helical phase fronts and quantized topological charge~$\ell$ possess a spatially nonlocal OAM density that cannot be defined at a single point \cite{ref9, ref10}, making them well-suited for coupling to higher-order multipolar modes.  These structured beams provide a means to explore chiral responses beyond the reach of conventional spin-based probes.  This raises a fundamental and previously unresolved question: Can the OAM of light generate a distinct form of optical chirality, referred to as orbital OC, that is capable of driving higher-order chiral responses?

This question has been explored under the concept of vortex or helical dichroism (VD or HD), which refers to the differential interaction of chiral matter with linearly polarized OVs depending on the sign of their OAM.  The use of linear polarization is intended to suppress spin OC contributions and isolate OAM-induced chirality.  VD has been reported in a wide range of systems, including chiral molecules in solution or powder form \cite{ref11, ref12, ref13, ref14}, plasmonic composites such as metal nanoparticle assemblies functionalized with chiral molecules \cite{ref15}, and chiral nanoparticles \cite{ref16, ref17} and microstructures \cite{ref18}.  However, several obstacles, on the optical, measurement, and material sides, have hindered unambiguous identification of orbital-specific chiral responses.  On the optical side, tightly focused linearly polarized OVs generate longitudinal field components, resulting in nonzero spin OC.  This spin OC, unlike that of CP light, does not reverse under mirror reflection, and can therefore mimic chiral signals despite not reflecting the intrinsic chirality of the OAM beam.  Experiments have shown that VD spectra can closely resemble CD spectra, suggesting a possible contribution from focusing-induced spin OC rather than orbital OC \cite{ref16, ref19}.  Similar trends are also seen in simulations in \cite{ref17}.  On the measurement side, the lack of systematic checks for reciprocity, a direct experimental signature of time-reversal symmetry, in many previous studies has led to ambiguity in identifying true chiral responses.  Most reports did not confirm reciprocity by reversing the illumination direction or sample orientation.  This omission is particularly concerning for anisotropic systems, where nonreciprocal signals can arise from extrinsic asymmetries and masquerade as chiral effects.  On the material side, many investigated structures are substantially larger than the wavelength\cite{ref13, ref15, ref18}, leading to excitation of multiple resonant modes and complicating attribution of the observed VD signals to specific multipolar interactions.  Despite recent refinements in experimental design, a conclusive identification of orbital OC remains elusive.  To unambiguously attribute observed chiral signals to orbital OC, it is essential to simultaneously address challenges on the optical, measurement, and material sides.

To overcome these limitations in a unified manner, we develop a comprehensive experimental strategy.  We employ twisted gold nanorod dimers (TNDs), subwavelength chiral nanostructures in which electric dipole and quadrupole resonances are spectrally well separated, enabling selective probing of higher-order chiral responses on a minimal and controllable platform.  As structured light, we use CP-OVs that carry both SAM and tunable OAM.  By holding the SAM constant while varying the OAM, quantified through~$\ell$ = 0, $+1$, $-1$, we isolate the contribution of orbital OC.  We define chiral dichroism (ChD) as a generalized framework encompassing both CD and VD, enabling unified exploration of spin- and orbital-induced chiral interactions.  To ensure robustness, we perform reciprocity checks by reversing the sample orientation during transmission measurements.  Spatially and spectrally resolved measurements on a single TND uncover distinct ChD signatures at the beam centre and periphery, revealing the spatially structured and multipolar character of the chiral light–matter interaction.  These signatures cannot be explained by spin OC alone and are accurately reproduced by full-field electromagnetic simulations.  The results provide direct evidence that OAM can drive chiral optical responses through orbital OC.  By establishing a concrete link between OAM and higher-order chiral responses, this work lays the foundation for a unified understanding of chiral light–matter interactions.  It opens a pathway for hierarchical chiral sensing across multiple length scales, including the mesoscopic regime, beyond the dipolar limit, and paves the way for future exploration of orbital OC in complex photonic systems.

\begin{figure}[H]
  \centering
  \includegraphics[width=88mm]{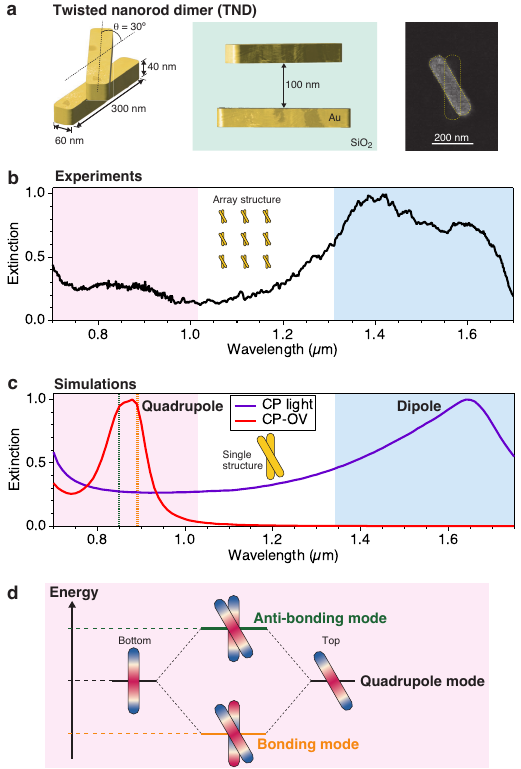}
  \caption{\textbf{Identification of electric quadrupole resonances in twisted gold nanorod dimers (TNDs).} \textbf{a,} Schematic of the TND geometry (left), consisting of two gold nanorods (300 nm $\times$ 60 nm $\times$ 40 nm) arranged with a 30$^{\circ}$ in-plane twist and a 100 nm vertical gap, embedded in a uniform SiO$_2$ matrix (middle), with a scanning electron microscopy (SEM) image of a fabricated TND (right). \textbf{b,} Measured extinction spectrum from a periodic TND array (500 nm pitch) under unpolarized plane-wave illumination. A dominant dipolar resonance appears near 1.5 $\mu$m, with a weaker shoulder around 850 nm attributed to higher-order modes. \textbf{c,} Simulated extinction spectra of a single TND under tightly focused circularly polarized (CP) light and CP optical vortex (CP-OV), showing that only CP-OV excitation activates a doublet near 850 nm, consistent with electric quadrupole resonances. \textbf{d,} Energy-level diagram illustrating bonding and anti-bonding electric quadrupole modes identified from simulated surface charge distributions. These hybridized modes arise from near-field coupling between the two nanorods and account for the spectrally resolved doublet in \textbf{c}.}
  \label{fig:TND}
\end{figure}

To isolate and verify the mechanism of ChD beyond the dipolar regime, we engineered a minimal and structurally well-defined chiral platform based on TNDs (Fig.~\ref{fig:TND}a). Unlike complex molecular or macroscopic systems \cite{ref11, ref12, ref13, ref14, ref15, ref16, ref17, ref18}, TNDs offer a geometrically controllable and symmetry-broken configuration that allows direct identification of the physical origin of ChD signals, particularly those mediated by orbital OC. Rather than treating TNDs as a typical example of a chiral sample, we use them as calibrated chiral probes, enabling systematic investigation of orbital OC--induced multipolar chiral light--matter interactions. Each TND comprises two gold nanorods with well-defined dimensions: 300~nm in length, 60~nm in width, and 40~nm in thickness. The rods are separated vertically by a 100~nm gap and arranged with a 30$^{\circ}$ in-plane angular twist. They are embedded in a uniform SiO$_2$ matrix, as confirmed by scanning electron microscopy (right panel of Fig.~\ref{fig:TND}a). This twist breaks inversion symmetry, imparting geometric chirality to the structure. The silica environment suppresses substrate-induced artifacts and ensures dielectric homogeneity, allowing the measured optical response to reflect the intrinsic multipolar properties of the TNDs.

To identify candidate resonances for ChD measurements, we first measured extinction spectra from a periodic array of TNDs with a 500~nm lattice period under unpolarized plane-wave illumination (Fig.~\ref{fig:TND}b). The spectrum exhibits a strong peak near 1.5~$\mu$m, corresponding to the bright electric dipole resonance, and a weaker spectral shoulder around 850~nm. However, the low contrast and spectral overlap in this region make it difficult to conclusively assign the 850~nm feature based on this measurement alone. This ambiguity arises because electric quadrupole modes are generally dark under uniform plane-wave excitation and do not efficiently couple to far-field radiation. The weak residual signal near 850~nm is likely due to collective effects in the periodic array, such as near-field coupling and diffraction-mediated excitation from neighboring TNDs, which can partially activate these otherwise dark modes. To clarify the origin of this spectral feature and directly assess the role of structured light, we performed full-field electromagnetic simulations using COMSOL Multiphysics to model the optical response of a single TND under tightly focused excitation by CP light and CP-OV (Fig.~\ref{fig:TND}c). The input beams were initially defined in the paraxial regime with SAM--OAM indices $(s,\,l)=(\pm1,\,0)$ for CP light and $(\pm1,\,\pm1)$ for CP-OV, and subsequently propagated through a high-NA (0.95) objective to generate the focused fields used for illumination. Under focused CP light illumination, the simulated extinction spectrum exhibits a single long-wavelength resonance consistent with electric dipole excitation, and no discernible response near 850~nm. In contrast, focused CP-OV illumination induces a distinct doublet near 850~nm, which is absent in the CP case, clearly indicating the activation of quadrupole resonances by structured light. This confirms that the shoulder feature in the experimental spectrum originates from electric quadrupole modes, which become accessible only through spatially structured excitation fields. To further characterize these resonances, we extracted the local electric field normal to the nanorod surfaces and estimated the surface charge distribution using Gauss's law. As shown in Fig.~\ref{fig:TND}d, the lower- and higher-energy peaks correspond to bonding and antibonding electric quadrupole modes, respectively. These arise from near-field coupling between quadrupole oscillations in the two rods, resulting in hybridized modes with distinct charge symmetries. The close agreement between simulated spectra and experimental extinction features validates the use of TNDs as a calibrated platform for isolating quadrupole-driven ChD. These findings establish the spectral fingerprint of electric quadrupole modes in TNDs and guide the selection of the 750--900~nm range for subsequent ChD measurements.

To quantitatively probe the intrinsically weak chiral optical response of a single TND, we developed a confocal optical microscopy system that achieves high spectral sensitivity and diffraction-limited spatial resolution. ChD signals are typically several orders of magnitude weaker than extinction or absorption, making single-nanostructure measurements challenging. To address this, we implemented a high-frequency SAM--OAM modulation scheme in which the excitation beam is rapidly switched between LH and RH CP-OV states at 50~kHz \cite{ref20}. Previous implementations based on spatial light modulators were limited to OAM switching rates below 100~Hz, within the low-frequency noise band, resulting in reduced measurement sensitivity \cite{ref13, ref14, ref15, ref16, ref17, ref18}. The resulting differential signal is selectively extracted via lock-in detection at the same frequency, effectively suppressing low-frequency noise and instrumental drift. By integrating a wavelength-tunable light source and confocal detection, the system enables spectrally and spatially resolved ChD measurements at the single-particle level.

\begin{figure}[H]
  \centering
  \includegraphics[width=\textwidth]{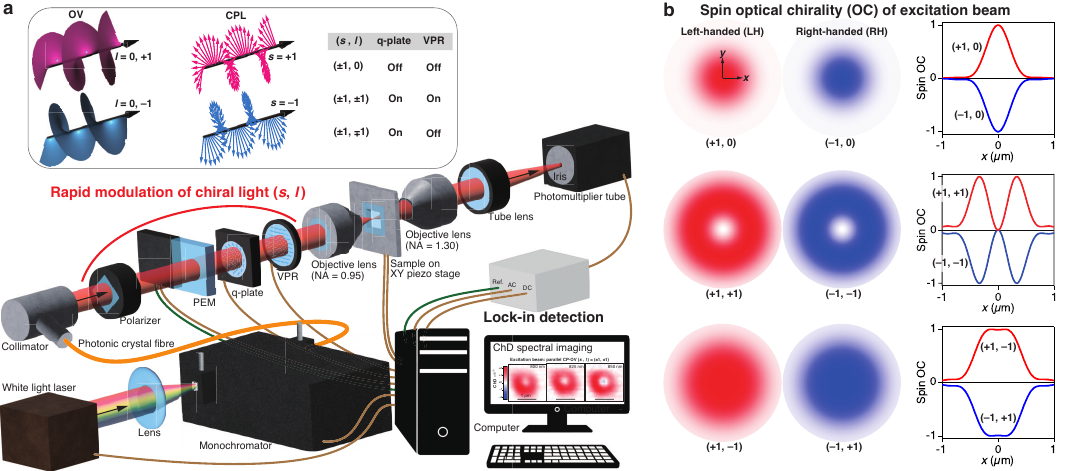}
  \caption{\textbf{High-sensitivity confocal microscopy system for spatially and spectrally resolved chiral dichroism (ChD) measurements.} \textbf{a,} Schematic of the experimental setup for detecting ChD from single twisted nanorod dimers (TNDs). The excitation beam is generated from a broadband laser filtered by a monochromator and modulated in spin angular momentum (SAM, $s$) using a photoelastic modulator (PEM). A q-plate converts SAM modulation into simultaneous SAM--orbital angular momentum (OAM, $l$) modulation to produce structured beams, and a variable phase retarder (VPR) controls the relative sign of SAM and OAM, enabling access to parallel ($s \times l > 0$) and anti-parallel ($s \times l < 0$) SAM--OAM configurations. The system generates three distinct beam states: standard CP light $(s,\,l) = (\pm1,\,0)$, parallel CP-OV $(\pm1,\,\pm1)$, and anti-parallel CP-OV $(\pm1,\,\mp1)$, as illustrated in the inset. The beam is tightly focused onto a single TND, and transmitted light is collected in a confocal geometry and demodulated via lock-in detection at 50~kHz. \textbf{b,} Numerically calculated spin optical chirality (spin OC) distributions in the focal plane for the three excitation beams. The CP light exhibits a uniform spin OC distribution peaked at the beam centre. The parallel CP-OV forms a ring-like spin OC profile with a central node, while the anti-parallel CP-OV yields a flattened central plateau due to the longitudinal field component induced by tight focusing. All profiles are normalized to their respective extrema.}
  \label{fig:apparatus}
\end{figure}

Figure~\ref{fig:apparatus}a illustrates the optical configuration. A broadband white-light laser is filtered by a monochromator to generate a tunable excitation beam, which is coupled into a photonic crystal fiber and collimated. The beam is linearly polarized and passed through a photoelastic modulator (PEM), which modulates the SAM between LH and RH circular polarization at 50~kHz. This SAM modulation is transformed into modulation of both SAM and OAM by a q-plate, which generates CP-OV via SOC \cite{ref21}. A variable phase retarder (VPR) further adjusts the relative sign between SAM and OAM, enabling control over parallel ($s \times l > 0$) and anti-parallel ($s \times l < 0$) configurations. By coordinating the PEM, q-plate, and VPR, the system produces three distinct beam states: $(s,\,l) = (\pm 1,\,0)$, $(\pm 1,\,\pm 1)$, and $(\pm 1,\,\mp 1)$, as shown in the inset of Fig.~\ref{fig:apparatus}a. The beam is tightly focused through a high-NA objective (NA = 0.95), and transmitted light is collected by a second objective (NA = 1.30). An adjustable iris located at the conjugate image plane acts as a pinhole, implementing confocal detection and enhancing spatial resolution. The signal is detected by a photomultiplier tube and demodulated by a lock-in amplifier at the modulation frequency.

To characterize the excitation beams, we numerically calculated the spin OC distributions in the focal plane for each beam state, as shown in Fig.~\ref{fig:apparatus}b. The spin OC, $C$, is defined as~\cite{ref22}
\begin{equation}
C = \frac{\varepsilon_0}{2} \, \mathbf{E} \cdot (\nabla \times \mathbf{E}) 
    + \frac{1}{2\mu_0} \, \mathbf{B} \cdot (\nabla \times \mathbf{B}),
\label{eq:OC}
\end{equation}
where $\mathbf{E}$ and $\mathbf{B}$ are the local electric and magnetic fields, and $\varepsilon_0$ and $\mu_0$ are the vacuum permittivity and permeability, respectively.  For standard CP light with $(s,\,l) = (\pm 1,\,0)$, spin OC is uniformly distributed and peaks on the optical axis, reflecting a spatially homogeneous helicity profile. In the parallel CP-OV configuration with $(s,\,l) = (\pm 1,\,\pm 1)$, spin OC forms a ring-like distribution with a central node, indicative of the vortex-induced spatial inhomogeneity. In contrast, the anti-parallel configuration $(s,\,l) = (\pm 1,\,\mp 1)$ yields a plateau-like spin OC profile centred on the optical axis, reflecting a redistribution of helicity density under tight focusing. This central enhancement primarily originates from the longitudinal electric field component induced via focusing. Importantly, the sign of spin OC at any point is determined solely by the SAM ($s$), not by the OAM ($l$) \cite{ref23}. This means that all three excitation beams share the same handedness-defined chirality at a given SAM, regardless of their OAM value. Therefore, if the chiral optical response of the TNDs arises predominantly from dipolar interactions, the spatial distribution of the ChD signal is expected to follow that of spin OC. Any deviation from this correspondence would suggest the involvement of higher-order multipolar processes and nontrivial contributions from the beam topology, such as orbital OC. This insight motivates the need for spatially resolved ChD measurements capable of discriminating between spin- and orbital-induced chiral interactions.

\begin{figure}[H]
  \centering
  \includegraphics[width=\textwidth]{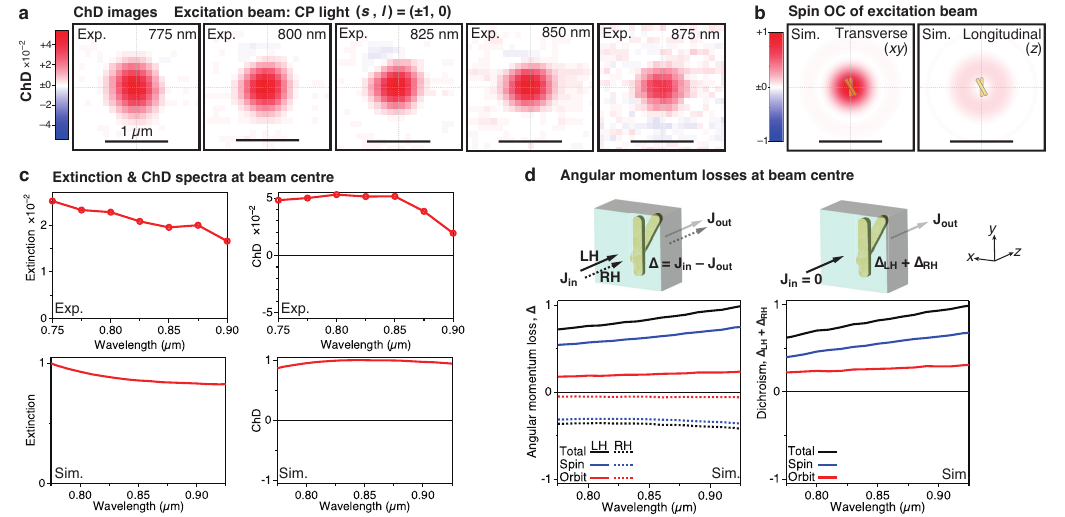}
  \caption{\textbf{Chiral dichroism (ChD) under circularly polarized (CP) light: spin optical chirality (OC)--dominated regime.} \textbf{a,} Spatially resolved ChD images of a single twisted gold nanorod dimer (TND) excited with CP light $(s,\,l) = (\pm 1,\,0)$, recorded at wavelengths spanning the quadrupole resonance range. The signal exhibits a symmetric, centre-peaked profile. \textbf{b,} Calculated spin OC distributions in the focal plane of the excitation beam, separated into transverse ($xy$) and longitudinal ($z$) components. Both maps are normalized using the larger of the two maximum values, allowing their relative magnitudes to be directly compared. \textbf{c,} Experimental (top) and simulated (bottom) extinction (left) and ChD (right) spectra at the beam centre. The ChD signal remains positive and featureless across the spectral range. Setting the twist angle to $\theta = 0^{\circ}$ of TND eliminates the ChD signal (Supplementary Materials), confirming its origin in structural chirality. \textbf{d,} Simulated angular momentum losses projected along the beam propagation ($z$-) axis ($\Delta = J_{\mathrm{in}} - J_{\mathrm{out}}$, left) and the resulting dichroism ($\Delta_{\mathrm{LH}} + \Delta_{\mathrm{RH}}$, right), separated into spin (blue), orbital (red), and total (black) components for left- (LH) and right-handed (RH) CP light excitation. The dichroism spectrum is dominated by the SAM contribution, with OAM remaining smaller but non-negligible. For the achiral configuration ($\theta = 0^{\circ}$), this analysis yields zero dichroism, as expected for a purely chiral response (Supplementary Materials). Simulated values in \textbf{c} and \textbf{d} are normalized by their respective extremum values.}
  \label{fig:ChDCPL}
\end{figure}

We begin our investigation of ChD by probing the TND response under CP light excitation, corresponding to the spin-only configuration $(s,\,l) = (\pm 1,\,0)$. Figure~\ref{fig:ChDCPL}a presents spatially resolved ChD images acquired at representative wavelengths within the quadrupole resonance range. In all cases, the ChD signal exhibits a symmetric, centrally peaked distribution that closely matches the spatial profile of the spin optical chirality (spin OC) calculated from only the transverse field components of the excitation beam (left panel of Fig.~\ref{fig:ChDCPL}b). Under CP illumination, the total spin OC is dominated by its transverse component, with negligible longitudinal contribution, resulting in a spatially uniform helicity distribution. The observed ChD therefore reflects spin OC--driven interactions, indicating that the TND response remains within the dipolar regime in this configuration.

To further examine the spectral characteristics, we extracted the ChD and extinction spectra at the beam centre, as shown in Fig.~\ref{fig:ChDCPL}c. Both quantities exhibit monotonic behavior across the measured range, with no pronounced peaks or dispersive features. The extinction spectrum gradually decreases with increasing wavelength, while the ChD remains positive and exhibits a broad plateau near 800~nm. Importantly, when the structural twist angle is set to $\theta = 0^{\circ}$, eliminating the geometric chirality of the TND, the ChD signal vanishes entirely (Supplementary Materials), confirming that the observed signal originates from the chiral asymmetry of the structure. Electromagnetic simulations under identical conditions reproduce these trends qualitatively, confirming that CP light primarily excites electric dipole modes in the TND and induces weak ChD responses mediated by spin OC. The absence of a distinct resonance feature further indicates that quadrupole modes, which are inaccessible under CP illumination, do not contribute significantly in this case.

To gain deeper insight into the underlying mechanisms, we analyzed the angular momentum ($J$) exchange between light and matter ($\Delta = J_{\mathrm{in}} - J_{\mathrm{out}}$) at the beam centre using a recently developed method~\cite{ref24} that enables separate and quantitative evaluation of SAM ($\Delta_{\mathrm{SAM}}$) and OAM ($\Delta_{\mathrm{OAM}}$) dissipation. This approach evaluates the angular momentum projected along the light propagation ($z$-) axis, excluding transverse components such as those from longitudinal fields, and decomposes the total loss into spin and orbital contributions ($\Delta = \Delta_{\mathrm{SAM}} + \Delta_{\mathrm{OAM}}$), even in the presence of complex scattering and near-field effects. Applied to the present system under focused CP light illumination, the simulations (Fig.~\ref{fig:ChDCPL}d) show that LH CP light experiences greater angular momentum loss than its RH counterpart ($\Delta_{\mathrm{LH}} > \Delta_{\mathrm{RH}}$), with the resulting dichroism ($\Delta_{\mathrm{LH}} + \Delta_{\mathrm{RH}}$) dominated by the SAM term while the OAM term remains smaller but non-negligible ($\Delta_{\mathrm{SAM}} > \Delta_{\mathrm{OAM}}$) across the spectrum. As a stringent check, applying the same analysis to the achiral configuration ($\theta = 0^{\circ}$) yields zero dichroism ($\Delta_{\mathrm{LH}} = -\Delta_{\mathrm{RH}}$), as expected for a purely chiral signal (Supplementary Materials). These results confirm that the measured ChD in this configuration is governed by spin OC--driven processes, with no significant orbital OC contribution under CP light excitation, and provide a rigorous basis for interpreting ChD in more complex beam configurations where both contributions may arise.

\begin{figure}[H]
  \centering
  \includegraphics[width=\textwidth]{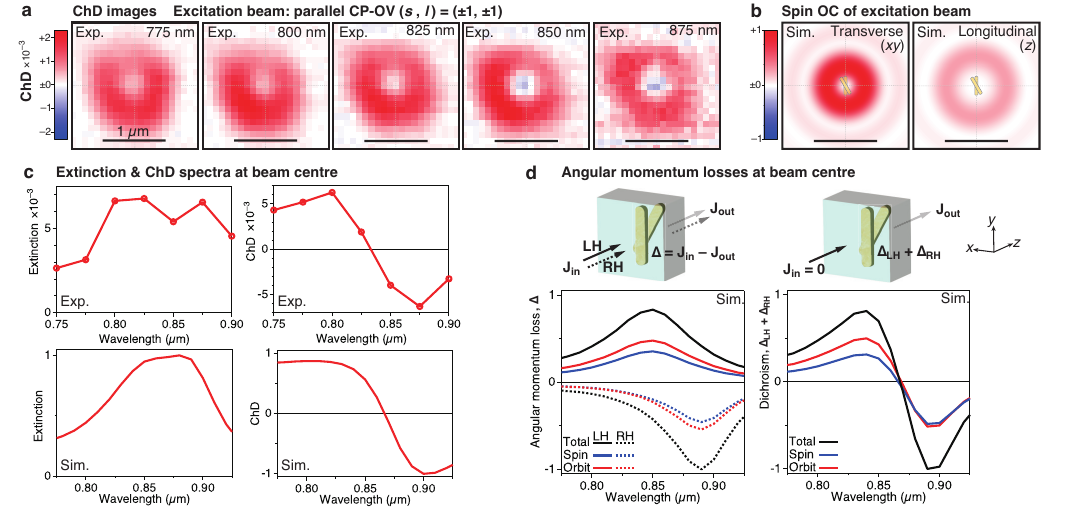}
  \caption{\textbf{Chiral dichroism (ChD) under parallel circularly polarized optical vortex (CP-OV) illumination: orbital optical chirality (OC)--dominated regime.}
\textbf{a,} Spatially resolved ChD images of a single twisted gold nanorod dimer (TND) under parallel CP-OV excitation with $(s,\,l) = (\pm 1,\,\pm 1)$, recorded at wavelengths spanning the quadrupole resonance range. Donut-shaped ChD profiles emerge near the quadrupole resonances, accompanied by a clear central response. 
\textbf{b,} Calculated spin OC distributions in the focal plane of the excitation beam, separated into transverse ($xy$) and longitudinal ($z$) components. Both maps are normalized using the larger of the two maximum values, allowing their relative magnitudes to be directly compared. 
\textbf{c,} Experimental (top) and simulated (bottom) extinction (left) and ChD (right) spectra at the beam centre. Distinct bonding and antibonding quadrupole resonances are observed, with the ChD exhibiting spectral sign inversion between the peaks. The ChD signal disappears when $\theta = 0^{\circ}$, confirming its origin in structural chirality. 
\textbf{d,} Simulated angular momentum losses projected along the beam propagation ($z$-)axis ($\Delta = J_{\mathrm{in}} - J_{\mathrm{out}}$, left) and the resulting dichroism ($\Delta_{\mathrm{LH}} + \Delta_{\mathrm{RH}}$, right), separated into spin (blue), orbital (red), and total (black) components for left- (LH) and right-handed (RH) parallel CP-OV excitation. Orbital contributions dominate over spin across the spectral range, in contrast to the CP light case (Fig.~\ref{fig:ChDCPL}d), evidencing a shift to an orbital OC--mediated interaction regime. Simulated values in \textbf{c} and \textbf{d} are normalized by their respective extremum values.
}
  \label{fig:ChDpCPOV}
\end{figure}

Figure~\ref{fig:ChDpCPOV} presents the central result of this work, demonstrating for the first time the quadrupole-mediated ChD of a single chiral nanostructure in a clear and spatially resolved fashion. We investigated the ChD response of the TND under excitation by parallel CP-OV beams with $(s,\,l) = (\pm 1,\,\pm 1)$, where the SAM and OAM are aligned. Figure~\ref{fig:ChDpCPOV}a shows the spatial ChD map at the quadrupole resonance wavelengths. A characteristic donut-shaped distribution is observed, with positive ChD signals localized in a ring surrounding the optical axis. This ring structure closely resembles the distribution of spin OC considering only transverse field components of the excitation beam shown in the left panel of Fig.~\ref{fig:ChDpCPOV}b, indicating that the observed ChD in this peripheral region is governed primarily by the chirality of SAM.

However, in strong contrast to spin OC expectations, a distinct and finite ChD signal is also observed at the beam centre. This centre corresponds to the phase singularity of the OV, where the electric field vanishes and spin OC is zero. Nevertheless, a clear ChD signal is detected, and its spectral characteristics are notably different from those of the surrounding ring. As shown in Fig.~\ref{fig:ChDpCPOV}c, the extinction spectrum at the beam centre exhibits two distinct peaks, which are consistent with the bonding and anti-bonding electric quadrupole resonances identified in simulations (Figure~\ref{fig:TND}c,d). These peaks appear more clearly than in the plane-wave excitation case shown in Fig.~\ref{fig:TND}b, indicating that the structured excitation beam enhances the quadrupole coupling by enabling selective mode excitation. The corresponding ChD spectrum displays a sign reversal between the two resonances. This spectral behavior mirrors that described in the classical Born--Kuhn model, where bonding and antibonding modes produce oppositely signed chiral responses~\cite{ref25}. Notably, when the structural twist angle is set to $\theta = 0^{\circ}$, rendering the nanorod dimer achiral, the ChD signal vanishes across the entire spectral range (Supplementary Materials), confirming that the observed chiral response originates from the geometric chirality of the structure.

To further understand this phenomenon, we analyzed the angular momentum loss projected along the beam propagation ($z$-) direction, as shown in Fig.~\ref{fig:ChDpCPOV}d. The dichroism ($\Delta_{\mathrm{LH}} + \Delta_{\mathrm{RH}}$) shows two distinct peaks, corresponding to the bonding and antibonding modes of quadrupole resonances. Notably, the peak at shorter wavelengths (anti-bonding mode) is more pronounced under LH CP-OV excitation, whereas the longer-wavelength peak (bonding mode) is stronger under RH CP-OV excitation. This peak shift reflects the structural chirality of the TND and the selective coupling of angular momentum components. Importantly, across the entire spectral range, the dissipation of OAM consistently exceeds that of SAM ($\Delta_{\mathrm{OAM}} > \Delta_{\mathrm{SAM}}$), in stark contrast to the CP excitation case shown in Fig.~\ref{fig:ChDCPL}d where the opposite relation ($\Delta_{\mathrm{SAM}} > \Delta_{\mathrm{OAM}}$) was observed. The ChD spectrum closely follows the dichroism ($\Delta_{\mathrm{LH}} + \Delta_{\mathrm{RH}}$), supporting the interpretation that the observed chiral response is governed predominantly by the OAM of light, or orbital OC.

Altogether, the spatially resolved imaging, the peak inversion between bonding and antibonding modes in the ChD spectrum, and the mode-selective angular momentum dissipation provide direct evidence that quadrupole-mediated ChD is governed by a previously unexplored regime of chiral light--matter interaction, where orbital OC plays a dominant role over spin OC. The emergence of a clear chiral response at the vortex centre, where both electric field and spin OC vanish, underscores the distinct contribution of orbital OC as a higher-order mechanism enabled by the structured phase topology of light.

\begin{figure}[H]
  \centering
  \includegraphics[width=88mm]{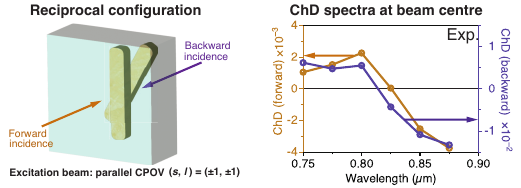}
  \caption{\textbf{Reciprocity test to confirm the chiral origin of the observed chiral dichroism (ChD).}
\textbf{Left,} Schematic of the reciprocal measurement configuration, in which the twisted gold nanorod dimer (TND) is illuminated by parallel circularly polarized optical vortex (CP-OV) from either the top (forward incidence) or bottom (backward incidence) side of the substrate. Both excitation and collection paths employ objective lenses with matching numerical apertures (NA = 0.95), ensuring symmetric optical conditions. 
\textbf{Right,} Experimentally obtained ChD spectra at the beam centre under forward (orange) and backward (purple) incidence. Despite minor variations in signal amplitude, the peak positions and spectral features are consistent between the two configurations, demonstrating the reciprocal nature of the response. This confirms that the ChD signals originate from the intrinsic geometric chirality of the TND, not from structural anisotropy or alignment-induced artifacts.}
  \label{fig:Reciprocity}
\end{figure}

Figure~\ref{fig:Reciprocity} presents the results of the reciprocity test, which serves as a definitive check for the chiral origin of the observed ChD signals. In this measurement, we replaced the collection-side objective lens with one having a NA of 0.95, matching that of the illumination-side objective. This symmetric optical configuration ensures that the numerical apertures for both excitation and detection are identical, thereby enabling a more rigorous verification of optical reciprocity. As established in the foundational experimental work~\cite{ref26,ref27}, and in Barron's theoretical framework of true chirality~\cite{ref1}, a genuine chiral optical response must remain invariant under the combined operations of time reversal and spatial inversion. In this context, spatial inversion refers to reversing the sample orientation with respect to the incident light.

The left panel of Fig.~\ref{fig:Reciprocity} shows a schematic of the experimental geometry, in which the TND is illuminated from either the top or the bottom side. In the right panel of Fig.~\ref{fig:Reciprocity}, we plot the ChD spectra obtained at the beam centre of parallel CP-OV beams for both illumination directions. While the overall magnitude of the signal varies slightly due to minor experimental differences such as focusing and alignment, the spectral features, including peak positions and relative signs, remain highly consistent across both measurements. This observation confirms that the recorded ChD signals are reciprocal and therefore not attributable to structural anisotropy or sample misalignment. Importantly, although the TND is an anisotropic structure, anisotropic scatterers can in general give rise to nonreciprocal signals that mimic chiral effects. However, the clear spectral agreement observed in this test demonstrates that the ChD response originates from the intrinsic geometric chirality of the structure rather than extrinsic asymmetries. This reciprocity result provides strong evidence that the multipolar ChD signal, particularly the quadrupole-driven response observed in Figure~\ref{fig:ChDpCPOV}, represents a true chiral optical interaction.

\begin{figure}[H]
  \centering
  \includegraphics[width=\textwidth]{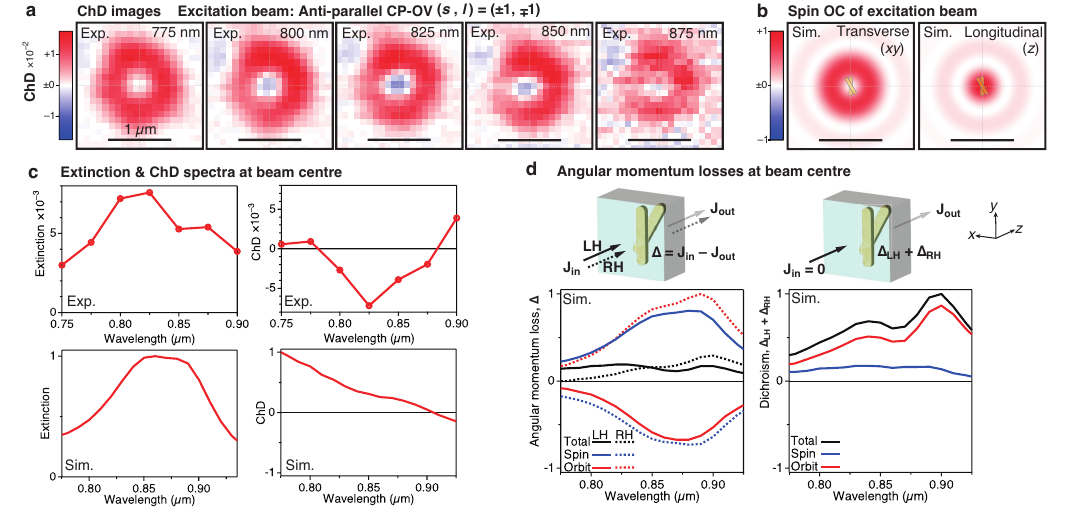}
  \caption{\textbf{Chiral dichroism (ChD) under anti-parallel circularly polarized optical vortex (CP-OV) illumination: Orbital angular momentum (OAM) sign sensitivity and orbital optical chirality (OC) dominance.}
\textbf{a,} Spatially resolved ChD images of a single twisted gold nanorod dimer (TND) under CP-OV excitation with anti-parallel spin--orbit configuration $(s,\,l) = (\pm 1,\,\mp 1)$, recorded at wavelengths spanning the quadrupole resonance range. The donut-shaped profile resembles that of the parallel configuration (Fig.~\ref{fig:ChDpCPOV}a). 
\textbf{b,} Calculated spin OC distributions in the focal plane of the excitation beam, separated into transverse ($xy$) and longitudinal ($z$) components. Both maps are normalized using the larger of the two maximum values, allowing their relative magnitudes to be directly compared. 
\textbf{c,} Experimental (top) and simulated (bottom) extinction (left) and ChD (right) spectra at the beam centre. The extinction shows bonding and antibonding quadrupole resonances, with a stronger antibonding response due to radial electric field alignment with the spatial charge distribution of the mode. The ChD signal reverses sign compared to the parallel case, evidencing OAM sensitivity. The signal vanishes for $\theta = 0^{\circ}$, confirming a chiral origin. 
\textbf{d,} Simulated angular momentum losses projected along the beam propagation ($z$-)axis ($\Delta = J_{\mathrm{in}} - J_{\mathrm{out}}$, left) and the resulting dichroism ($\Delta_{\mathrm{LH}} + \Delta_{\mathrm{RH}}$, right), separated into spin (blue), orbital (red), and total (black) components for left- (LH) and right-handed (RH) anti-parallel CP-OV excitation. Orbital contributions in the dichroism exceed spin across the spectrum. Simulated values in \textbf{c} and \textbf{d} are normalized by their respective extremum values.
}
  \label{fig:ChDapCPOV}
\end{figure}

Figure~\ref{fig:ChDapCPOV} investigates how the ChD response changes when the sign of the OAM of the excitation beam is reversed, while keeping the SAM unchanged. This anti-parallel configuration, defined by $(s,\,l) = (\pm 1,\,\mp 1)$, provides a direct means of distinguishing between spin-mediated and orbitally mediated contributions to the observed chiral signal. If the ChD response is sensitive to the sign of OAM, it would strongly suggest that the interaction is governed by the orbital component of the optical field rather than its spin component.

As shown in Fig.~\ref{fig:ChDapCPOV}a, the measured ChD images under anti-parallel CP-OV illumination display a donut-like spatial profile, which is qualitatively similar to that observed in the parallel configuration shown in Fig.~\ref{fig:ChDpCPOV}. This distribution closely resembles the spin OC profile calculated using only transverse field components (left panel of Fig.~\ref{fig:ChDapCPOV}b), but shows little resemblance to the spin OC profile based on longitudinal fields (right panel of Fig.~\ref{fig:ChDapCPOV}b). This discrepancy likely arises from the experimental geometry: in our confocal detection scheme, signals are collected in the image plane, which is insensitive to the longitudinal ($z$-polarized) field components.

Figure~\ref{fig:ChDapCPOV}c presents the extinction and ChD spectra obtained at the beam centre. As in the parallel configuration, two peaks associated with bonding and antibonding quadrupolar modes are observed, but here the antibonding mode exhibits stronger extinction. This can be understood from the electric field distribution of the excitation beam, which radiates outward from the beam centre and thus preferentially couples to the antibonding mode characterized by radial charge separation (Supplementary Materials). Notably, these quadrupole peaks were not clearly resolved under plane-wave excitation (Figure~\ref{fig:TND}b), again highlighting the advantage of structured light in selectively exciting multipolar modes.

The corresponding ChD spectrum shows a sign inversion in the antibonding mode compared to the parallel case, indicating that the observed signal is sensitive to the sign of OAM. Moreover, the ChD signal disappears entirely when the structural twist angle is set to $\theta = 0^{\circ}$, where the nanorod dimer becomes achiral. This confirms that the observed response originates from the intrinsic geometric chirality of the structure. Although the simulated ChD spectrum does not fully reproduce the experimental behavior, this discrepancy is particularly noticeable in the anti-parallel case, where the longitudinal component of the spin OC is comparable to or even larger than its transverse component (Fig.~\ref{fig:ChDapCPOV}b). Because our confocal detection scheme is primarily sensitive to transverse fields in the image plane, the enhanced longitudinal contribution observed in this configuration, which is not seen in the other cases, is likely responsible for reducing the agreement between experiment and simulation. 

Nevertheless, the angular momentum loss analysis in Fig.~\ref{fig:ChDapCPOV}c provides further insight. It is important to note that the angular momentum loss is evaluated along the beam propagation ($z$-)direction, and does not include contributions from the $z$-polarized fields. The calculated dichroism in angular momentum dissipation spectra indicates that the OAM contribution clearly dominates over the SAM contribution. These results provide compelling evidence for the existence of orbital OC. The presence of a measurable ChD signal at the centre of the vortex beam, where the transverse field amplitude vanishes and spin OC is nearly suppressed in the CP-OV case, highlights the distinct roles of spin and orbital contributions and supports the interpretation that orbital OC plays a dominant role in the observed response.

In summary, we have demonstrated that a single TND, when excited by parallel CP-OV beams, exhibits clear and spatially resolved ChD responses that cannot be explained by spin OC alone. Most strikingly, a strong ChD signal appears at the beam centre, where the transverse field amplitude vanishes and spin OC is suppressed, providing direct evidence of a quadrupole-mediated chiral interaction governed by the OAM of light. This reveals a distinct form of field chirality, orbital OC, enabling higher-order chiral responses beyond the conventional dipolar regime.

Despite this experimental advance, a comprehensive theoretical framework that describes ChD involving multipolar processes remains undeveloped. Existing models such as the Tang--Cohen formalism~\cite{ref8} are restricted to electric--magnetic dipolar interference and cannot describe quadrupolar or higher-order contributions. Developing a general theory of ChD that incorporates both SAM and OAM, as well as multipolar transitions, is therefore a key challenge.

Our results also shed light on the physical origins of enhanced chiral light--matter interactions in structured fields. In landmark work, Hendry \textit{et al.} showed that chiral molecules on chiral plasmonic nanostructures exhibit optical activity enhanced by up to six orders of magnitude~\cite{ref28}. While such enhancement has been partly attributed to spin OC amplification in the dipolar regime, this mechanism alone cannot fully account for the magnitude. They proposed that strong near-field gradients activate higher-order multipolar transitions, particularly electric quadrupole excitations, as originally formulated by Efrima~\cite{ref29}. In light of our findings, which directly demonstrate quadrupole-mediated ChD driven primarily by orbital OC using structured beams, we suggest that the same mechanism may underlie these plasmon-enhanced chiral effects. In both cases, strong field gradients and topological phase structures enable access to multipolar chiral transitions, pointing to a unified picture in which orbital OC is the common driver.

By establishing orbital OC-induced ChD at the single-nanostructure level, this work introduces a new paradigm for probing complex chiral phenomena and underscores the broader importance of structured light in chiral photonics. Future efforts should aim to build a rigorous multipolar theory of ChD and to explore orbital OC across diverse materials and excitation schemes. Such developments could pave the way toward scalable, high-resolution, and symmetry-selective chiral spectroscopy.

\begin{acknowledgments}
This work was supported by Grants-in-Aid for Scientific Research (KAKENHI) (Nos. JP24H00424, JP22H05132 in Transformative Research Areas (A) “Chiral materials science pioneered by the helicity of light” to Y.Y.T., and Nos. JP23K04669, JP21K14594 to S.H.) from the Japan Society for the Promotion of Science (JSPS), and JST FOREST Program (No. JPMJFR213O to Y.Y.T.).
\end{acknowledgments}

\bibliography{refs}

\end{document}